\documentclass[runningheads]{llncs}
\usepackage{graphicx}
\usepackage{xcolor}
\usepackage{subcaption}
\usepackage{array}
\usepackage{multirow}
\begin{document}
\title{Introducing CALMED: Multimodal Annotated Dataset for Emotion Detection in Children with Autism\thanks{Supported by Science Foundation Ireland under Grant number SFI/12/RC/2289\_P2.}}
\titlerunning{CALMED: Multimodal Emotion Detection Dataset for ASD}
%
\author{Annanda Sousa\inst{1}\orcidID{0000-0002-0388-3641} \and
Karen Young\inst{1}\orcidID{0000-0002-9452-2293} \and
Mathieu d'Aquin\inst{2}\orcidID{0000-0001-7276-4702}
\and
Manel Zarrouk\inst{3}\orcidID{0000-0002-8160-5671}
\and
Jennifer Holloway\inst{4}\orcidID{0000-0001-9197-641X}}
\authorrunning{A. Sousa et al.}
%
\institute{University of Galway, Ireland \\
\email{\{a.defreitassousa1,karen.young\}@nuigalway.ie}\and
K Team, LORIA CNRS/INRIA/Université de Lorraine, Nancy, France
\email{\{mathieu.daquin\}@loria.fr}\\\and
LIPN, Université Sorbonne Paris Nord, France\\
\email{\{azarrouk\}@sorbonne-paris-nord.fr}
\and
ASK - All Special Kids, Geneva, Switzerland\\
\email{\{jennifer.holloway\}@allspecialkids.org}}
\maketitle              
\begin{abstract} 
Automatic Emotion Detection (ED) aims to build systems to identify users' emotions automatically. This field has the potential to enhance HCI, creating an individualised experience for the user. However, ED systems tend to perform poorly on people with Autism Spectrum Disorder (ASD). Hence, the need to create ED systems tailored to how people with autism express emotions. Previous works have created ED systems tailored for children with ASD but did not share the resulting dataset. Sharing annotated datasets is essential to enable the development of more advanced computer models for ED within the research community. In this paper, we describe our experience establishing a process to create a multimodal annotated dataset featuring children with a level 1 diagnosis of autism. In addition, we introduce CALMED (Children, Autism, Multimodal, Emotion, Detection), the resulting multimodal emotion detection dataset featuring children with autism aged 8-12. CALMED includes audio and video features extracted from recording files of study sessions with participants, together with annotations provided by their parents into four target classes. The generated dataset includes a total of 57,012 examples, with each example representing a time window of 200ms (0.2s). Our experience and methods described here, together with the dataset shared, aim to contribute to future research applications of affective computing in ASD, which has the potential to create systems to improve the lives of people with ASD. 

\end{abstract}
\section{Introduction}

Affective Computing is a relatively new area in Computer Science that aims to create computer systems able to identify, process, respond to and generate emotions in human users~\cite{calvo2015oxford,tao2005affective}. One of the most commonly investigated areas of Affective Computing is automatic Emotion Detection (ED)~\cite{calvo2015oxford}, also referred to as affect recognition, affect detection, and emotion recognition. ED aims to automatically identify people's cognitive states or emotions, e.g. happiness, anger, and fear~\cite{Liu2015}. ED systems, utilising different media inputs such as texts, video, audio and sensor signals, extract implicit cues from facial expression, eye gaze, and tone of voice. When combining more than one type of data, they are called \textit{Multimodal} Emotion Detection systems, which tend to outperform unimodal systems~\cite{Dmello2015,Soleymani2017}.

Most of the advancement in ED has been focused on the general population, i.e. users with typical neurological development, usually referred to as neurotypicals. When applying those systems to a specific population, e.g. children with autism, the systems usually do not perform well, mainly because of this particular population's way of expressing emotions~\cite{Liu2008}. Autism Spectrum Disorder (ASD) is a developmental disorder with a spectrum manifestation of traits characterised by impairments in social interaction, communication and repetitive patterns of behaviour and interests~\cite{Association2013}. Among the results of a recent meta-analysis~\cite{trevisan2018facial} that compared the facial expression production between a typical development (TD) population and people with autism, they found evidence that people with autism display facial expressions less often and less frequently than people with TD. Also, people with ASD expressions are lower in quality and less accurate. In the work of~\cite{Grossard2020}, the results showed that a Random Forest model needs more facial landmarks to classify facial expressions from children with autism than it needs from children with typical development. Those works together provide additional evidence that ED systems developed for children with typical development do not perform well when applied to children with autism, motivating the need to develop ED systems specifically tailored to children with autism. ``Level 1 diagnosis of autism'' is a terminology defined by DSM-5R~\cite{Association2013} referring to ASD without significant cognitive and language impairments. Also sometimes referred to as high-functioning autism, previously known as Asperger Syndrome~\cite{Gaus2011}. Throughout this paper, we will use the terms ASD and autism interchangeably.

Annotated datasets are a fundamental part of the creation of emotion detection systems. The emergence of more advanced computer models is heavily impacted by sharing these datasets within the research community. However, the creation and sharing of those datasets lead to a series of ethical challenges, requiring the research team to establish specific measures for protecting participants' rights, privacy, and well-being while maintaining the value of the created dataset to the research and the research community. Those issues appear from the first step of data collection, which often involves eliciting, capturing and tagging people's emotional expressions, to disseminating a sharable version of the data. When the emotion detection dataset features children with autism, representing a case of a vulnerable population with a medical condition, additional concerns must be considered. Examples of ethical matters in this scenario include: 
\begin{itemize}
    \item selecting tasks to evoke the target emotions without causing emotional harm to the children, 
    \item addressing the participant’s right to privacy as opposed to the important task of sharing resources in the research community, and 
    \item designing a data protection plan to comply with the participants’ rights as defined by the General Data Protection Regulation (GDPR).  
\end{itemize}
Previous research works have investigated the challenges of developing an emotion detection system tailored for children with autism.~\cite{Liu2008,Chu2018,Dawood2018,Kushki2015,Sarabadani2018}. These studies created ED systems for children with ASD,  demonstrating that it is viable to model how this population expresses emotions and automatically predicts their emotions. As part of their work, they created their own annotated datasets. In addition, the works of~\cite{Samad2018,ElKaliouby2007} specifically created annotated datasets featuring individuals with ASD. Samad's dataset~\cite{Samad2018} includes facial action units instead of emotions. Meanwhile, ElKaliouby's dataset~\cite{ElKaliouby2007}  contains data on adults with ASD instead of children. All these works reported having had to go through the phases of creating an emotion detection annotated dataset. However, the authors of those works did not share their resulting dataset nor the resources they generated during the dataset creation.

In this paper, we describe our experience establishing a process and measures to design a data collection framework with the ultimate goal of creating a multimodal emotion detection annotated dataset featuring children with a level 1 diagnosis of autism. 

We then introduce CALMED (Children, Autism, Multimodal, Emotion, Detection), the resulting multimodal (video-audio) emotion detection dataset. Children from 8-12 years old with a level 1 diagnosis of autism were invited to perform computer-based emotion elicitation tasks to evoke each of the target emotion zones. The dataset is annotated with four emotion zones (green, yellow, red and blue) based on the ``emotion zones for regulation'' framework~\cite{kuypers2013zones}. The annotation task was performed by the participants' parents, who watched and selected which emotion zone their child was expressing at each moment. 

The top four modalities of data input used in the emotion detection field are: 1. video, 2. physiological signals, 3. audio, and 4. text~\cite{saxena2020emotion}. Initially, CALMED was planned to include three modalities, video, audio and physiological signals. However, due to COVID-19 restrictions, the experiment setup required to be adapted and physiological signals input had to be excluded from the multimodal dataset since it would require close contact with participants to collect this data. Thus, our multimodal dataset includes two modalities of data input: video and audio. This bimodality is the most explored in multimodal emotion detection~\cite{calvo2015oxford}. 

CALMED, the annotated dataset for emotion detection described in this paper, has the following features:

\begin{itemize}
\item Compliant with Ethics guidelines and recommendations, with a main focus on the participants' emotional well-being;
\item Compliant with Data Protection legislation for handling personal data;
\item Multimodal dataset (video and audio data inputs);
\item Features children (8-12 years old) with a diagnosis of level 1 autism;
\item Annotated by the participants' parents into four classes representing emotion zones;
\item Data collected in a naturalistic setup, not posed emotions;
\item Data collected using computer-based task environment, as opposed to dialect eliciting tasks, i.e., from a conversation with the participant.
\end{itemize}

We also share the resources we created during the process, with the hope that they will benefit future research in creating and sharing their own emotion detection dataset featuring particular population, especially ASD population. Thus, future researchers can reuse the resources here presented, or customise them to adapt to their specific research goals.       
The additional resources shared in this work are three artefacts, namely:
\begin{itemize}
\item a computer-based task environment web system, for emotions' elicitation sessions;
\item an annotation web system, to support the dataset annotation process;
\item a working dataset generation system, to create an annotated dataset from the annotation process.
\end{itemize}

By sharing the dataset, this paper's contribution is unique since none of the previous research that created emotion detection datasets involving children with autism shared a dataset artefact with the research community, primarily due to privacy issues. While ED for the general population has plenty of available resources, e.g. annotated datasets, ED for ASD suffers from scarce available resources, with not even one annotated dataset available. Thus, the research team believes that making more resources available to support the creation and sharing of annotated datasets featuring populations with ASD can lead to advancing this specific application of the ED field. 

The rest of the paper is organised as follows, Section~\ref{sec:data-collection} describes the process and methods used to elicit and capture the original data from the participants. Then, Section~\ref{sec:dataset-creation} describes the methodology to create and process the multimodal annotated dataset. This is followed in Section~\ref{sec:analysis-discussion}, with a description of the dataset characteristics and a discussion about its applicability, limitations and perspectives. Finally, the paper concludes with a discussion of future work directions. 

\section{Data Collection} \label{sec:data-collection}
As introduced before, previous works~\cite{Samad2018,ElKaliouby2007,Liu2008,Chu2018,Dawood2018,Kushki2015,Sarabadani2018} have created affective computing systems focussing on people with ASD; each of these was required to create their own dataset since none was available.  In general, they followed the steps enumerated below. 

\begin{enumerate}
\item Modelling emotions definition,
\item define and design the eliciting tasks,
\item design the study session and task environment,
\item define data to be collected and create a data protection plan,
\item apply for ethics approval,
\item define inclusion and exclusion criteria of participation in the study,
\item recruit participants,
\item run study sessions,
\item define annotators, and running annotation sessions,
\item generate the final working annotated dataset from the original data collected. 
\end{enumerate}

This section describes the methods used while creating CALMED for steps 1-8, while Section \ref{sec:dataset-creation} describes the methodology for steps 9-10.

\subsection{Modelling Emotions}

Modelling and representing emotions is one of the inherent challenges in affective computing. Emotion definitions come from psychology and might be imprecise or fuzzy, making them difficult to apply to computer models~\cite{calvo2015oxford}. Usually, in affective computing, emotions are represented by the seven basic emotions, i.e. surprise, happiness, anger, disgust, contempt, sadness and fear, making them a set of categories. We can also find emotions represented according to the two dimensions of arousal (strong or weak) and valence (positive or negative), which usually take continuous value between -1 and 1~\cite{ringeval2013introducing}. 

However, for emotion detection applied for ASD, researchers tend not to work with the basic emotions from the general ED field. They argue that basic emotions are not the best target emotional states for applications of ED for autism from a pragmatic perspective. They, instead, work with different emotion states, e.g. anxiety (a prevalent co-occurring condition to ASD) or engagement level for instance~\cite{Liu2008}.

Following the previous related works, which did not focus on the seven basic emotions, , we selected other emotional states as labelling targets for CALMED dataset. We labelled the data using a framework called ``\textit{the zones of regulation}"~\cite{kuypers2013zones}. This framework is extensively used in psychology to help children with ASD, and other neurological conditions learn emotion regulation since it is common for children with ASD to present impairments in emotion regulation~\cite{scarpa2011improving}.

\begin{figure}[ht!]
\centering
\includegraphics[width=0.8\textwidth]{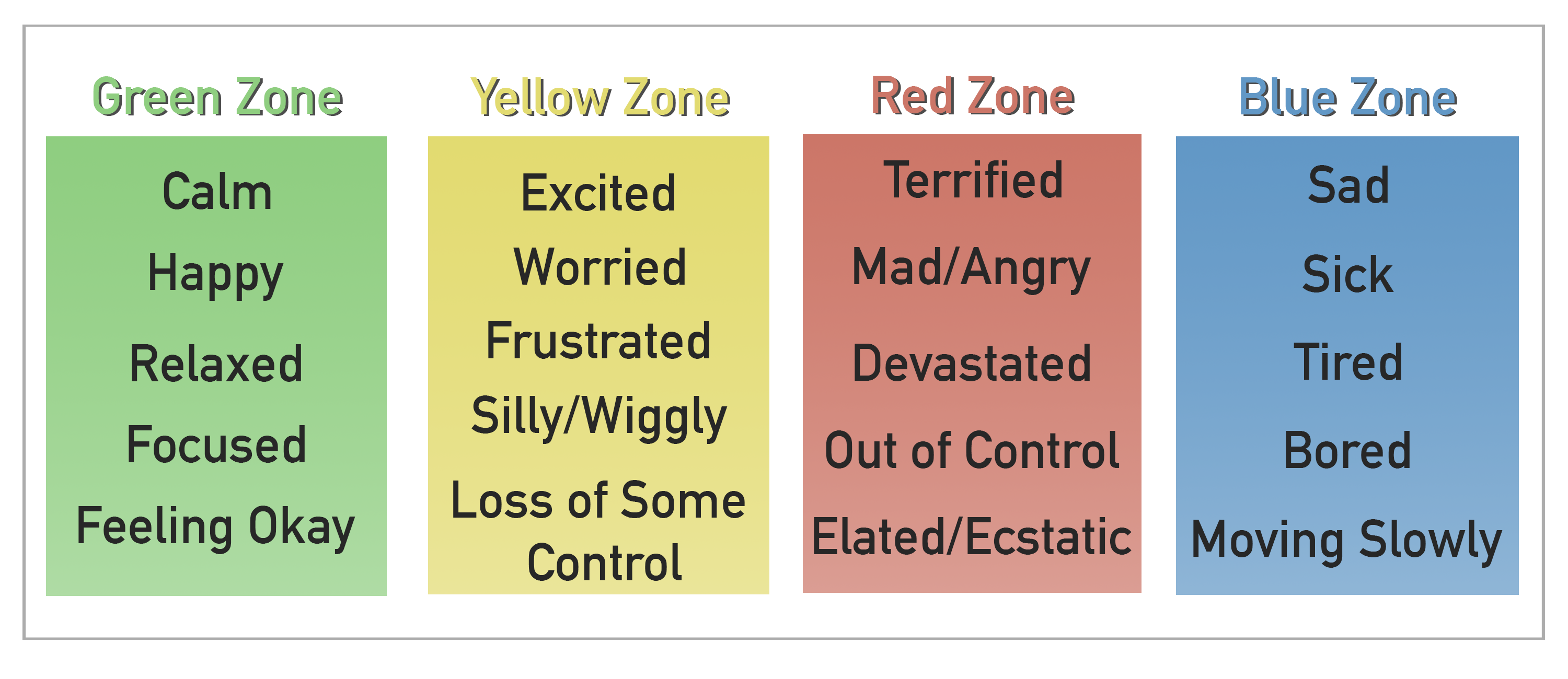}
\caption{The Four Emotion Zones, based on ``The Zones of Regulation", a Social Emotional Learning Framework \cite{kuypers2013zones}. The CALMED dataset is annotated with these four labels.} 
\label{fig:emotion-zones}
\end{figure}

\textit{The zones of regulation} framework has four different zones represented by colours (See Figure \ref{fig:emotion-zones}). One of the emotion zones is the calming zone, represented by the \textbf{green} zone. This ideal state is where the child is calm, relaxed, and ready to work, listen, and interact. The warning zone (\textbf{yellow} zone) indicates that the child is presenting signals of agitation or excitement. This state can originate from both positive and negative emotions. It can start from intense happiness or excitement and also from frustration. The high-agitation zone (\textbf{red} zone) indicates that the child is upset or angry, presenting severe difficulties in keeping control of their emotions. The last zone is the slowing zone, the \textbf{blue} zone, in which the child is on low energy and showing emotional signals of being sad, tired, sick or bored. In this state, the child might move slower than usual, stop speaking or show delays in interactive responses.

We obtain several benefits by using this emotions zones' framework as a target for labelling emotions. Firstly, the framework includes guidelines on activities to lead children back to the calming zone, and since the ultimate goal of creating a multimodal dataset is to develop affective computing systems to support children with ASD, using the "zones of regulation framework" makes it easy to incorporate the framework's activities to calm the child within an affect-sensitive interface. Second, parents of children with ASD are more likely to be familiarised with this framework because it is commonly used in the context of autism, hence making the labelling task by the parents more comfortable. Thirdly, considering the children's well-being, it is less harmful to the emotional comfort of children with ASD during the emotion elicitation experiment to elicit the four emotion zones than other strong negative emotions, e.g. fear and anxiety. This way we could elicit and observe all the four zones without causing a strong emotional discomfort on the participants. 

\subsection{Ethics}
This research is based in Ireland; thus, it must comply with Irish legislation and GDPR. The data collection involves children, which generates an additional challenge to address. Ireland has specific regulations for research involving children as participants. For instance, assent from the child is mandatory in addition to the parent's consent. Thus, if the parent allows the child to participate, but the child does not want to, the researcher can not go further with the data experiment. Such cases were encountered during this research's data collection. To obtain the assent from the child, we need to use age-appropriate language while being transparent and explaining what will be asked of them. Furthermore, we created age-appropriate materials to seek assent from children and to explain to them what the research participation would entail. We only talked to the children after obtaining permission from their parents. 

The most critical ethical concern is not to cause harm during the data collection sessions. The data collection objective was to capture naturalistic manifestations of emotions from children with autism. So we needed to elicit these emotions to record them and later annotate them. There are positive and negative emotions, so we decided early on not to elicit strong negative emotions which could cause emotional discomfort for the children, e.g. fear and phobia. 

We also put in place additional measures to protect the participants' emotional well-being. All sessions happened in the presence of a postgraduate psychology student who had experience working with children with ASD. This measure provided specialised support in case the participants got overwhelmed by emotions during the session and could not regulate themselves. As additional measures for emotional well-being, calming activities were included between each eliciting task to support the participants' emotion regulation, helping them calm themselves during the study session. 

The University of Galway's research ethics committee reviewed and approved the proposed ethics measures.  

\subsection{Data Protection}
Another important aspect of this data collection framework is the data protection plan. We needed to ensure that the personal data collected was secure and kept private from unauthorised access. Personal data is defined as any data that can be used to identify an individual, either by direct identification or through some sort of data engineering. To define the action and measures to put in place to protect the personal data and the participants' rights, we conducted a Data Protection Impact Assessment (DPIA), generating a plan with specific action points to follow on how to deal with the personal data resulting from this project. The university's data protection officer approved the DPIA. Some examples of actions we followed to ensure the participant's rights include: 
\begin{itemize}
    \item Provide comprehensive information on all aspects of the research for participants' parents, with a project's website, information sheets and other materials. 
    \item Provide information on the participants' right to withdraw from the research. 
    \item Provide information on the participants' right to have their data deleted.  
    \item Protect the right to privacy, ensuring only the research team can access the participant's data.
\end{itemize}
To make the information more available to participants and the general public, we created a website with all the documents and relevant information about the project.\footnote{\url{http://emotion-asd.datascienceinstitute.ie}}.

\subsection{Privacy and Sharing Resources}
None of the previous works which created annotated emotion detection datasets featuring children with autism shared their resulting dataset mainly due to privacy reasons. Since the raw recording files contain personal data, we needed to eliminate the identifiable aspect of the data to make it possible to share an annotated dataset. The approach we selected is, instead of creating a dataset featuring the raw video and audio files, to create a dataset with numerical and categorical features extracted from the original files. More detail on the feature generation can be found in the Sections~\ref{subsec:video-features} and~\ref{subsec:audio-features}. 

Making only available the extracted features instead of the original files is not an ideal approach to sharing resources in the research community because the researchers will not be able to extract their features. However, that is a necessary step, and the benefits of having a dataset with extracted features outweigh the previous limitation of not having a dataset at all. Not having an available dataset resulted in each new research work ought to start from scratch, creating a new dataset with all the steps and procedures involving its creation, consuming much time, material and people resources, which might be a limiting factor when defining a research project, leading to less research being conducted in this specific application.  

\subsection{Study Session Setup}

Our goal for the study session was to collect video and audio data from children with autism while they were engaged in the four emotion zones. Thus, we needed to elicit each emotion zone during a study session with participants. The elicitation is usually done by asking participants to perform evoking tasks. In order to create a more valuable dataset, it is essential to collect data following a setup as close as possible to the intended use-case scenario~\cite{ritter2012running}. For our case, we visioned a multimodal dataset as a resource to create affect-sensitive computer systems for children with ASD. Therefore, the closest scenario to elicit emotions is a computer-based task environment, i.e. where the participants interact with a computer system during the elicitation study session. 

The study sessions happened entirely online, over a Zoom video call, which was recorded. Initially, the study session was planned to happen in a lab room prepared with a computer, a high-resolution camera and a semi-professional microphone. However, due to pandemic restrictions, we could not conduct face-to-face meetings, making it necessary to adapt the study session to an online remote setup. The session was then a Zoom video call in which the researcher first talked to the participant, using age-appropriate language, seeking their assent and explaining the session and tasks. The participant then accessed an URL\footnote{\url{http://task-environment.datascienceinstitute.ie/}} to the task environment software and clicked the ``Start button'' to start the session. The system is programmed to automatically follow each task in sequence without requiring manual human commands. While the child is engaged with the task environment software, the Zoom video call is still on, recording the participant's camera and audio. 

The resulting video/audio file from the participant session is the raw file for the dataset creation. Using the participant's camera and microphone, on one side, does not guarantee the quality of the input, creating an additional challenge of noise and low-quality media. However, on the other side, it generates more realistic videos in a setup that an eventual affective computing system is more likely to encounter.  

We invited children and their parents to participate in the data collection study. The inclusion criteria were that the child had to have a level 1 diagnosis of ASD and to be within the age range (8-12 years old). We limited the age range to before the teenage phase, where other co-occurring conditions usually manifest and influence how the person expresses emotions. In addition, we needed to have written consent from the parent or guardian and verbal assent from the child. An exclusion criterion was having a history of cognitive or language impairment that does not fall under the level 1 diagnosis of autism and not attaining assent from the child. The study session was designed to last for no more than 30 minutes, and we asked each participant to attend two different sessions. 

As reported by all previous works, finding participants was challenging, especially because we were amidst a global pandemic. We had study sessions with four participants, one girl and three boys, with an average age of 10.25 (+-1.7). Table~\ref{tab:participants} summarises the participants' details and the number of sessions they attended. All the sessions generated close to four hours of recorded video/audio files.

\begin{table}[!ht]
    \centering
    \caption{Summary of participants information}
    \begin{tabular}{cccc}
    \hline
        \textbf{} & \textbf{Gender} & \textbf{\# Sessions} & \textbf{Age} \\ \hline
        \textbf{Participant 1} & Male & 1 & 8 \\ 
        \textbf{Participant 2} & Female & 2 & 12 \\ 
        \textbf{Participant 3} & Male & 2 & 10 \\ 
        \textbf{Participant 4} & Male & 2 & 11 \\ \hline
    \end{tabular}
    \label{tab:participants}
\end{table}

All study sessions followed the following agenda: 
\begin{enumerate}
\item Initial greetings with discussion on the child's interests (4 minutes).
\item Explanation of the session using age-appropriate language (3 minutes).
\item The session - the child accesses and follows the tasks on a web system (20 minutes).
\item Ending conversation (2 minutes).
\item Gifting the child with a certificate - with visual art involving the child's special interests (1 minute).
\end{enumerate}

\subsection{Emotion Elicitation Tasks}
\begin{figure*}[ht!]
    \centering
    \begin{subfigure}[t]{0.5\textwidth}
        \centering
        \includegraphics[width=\textwidth]{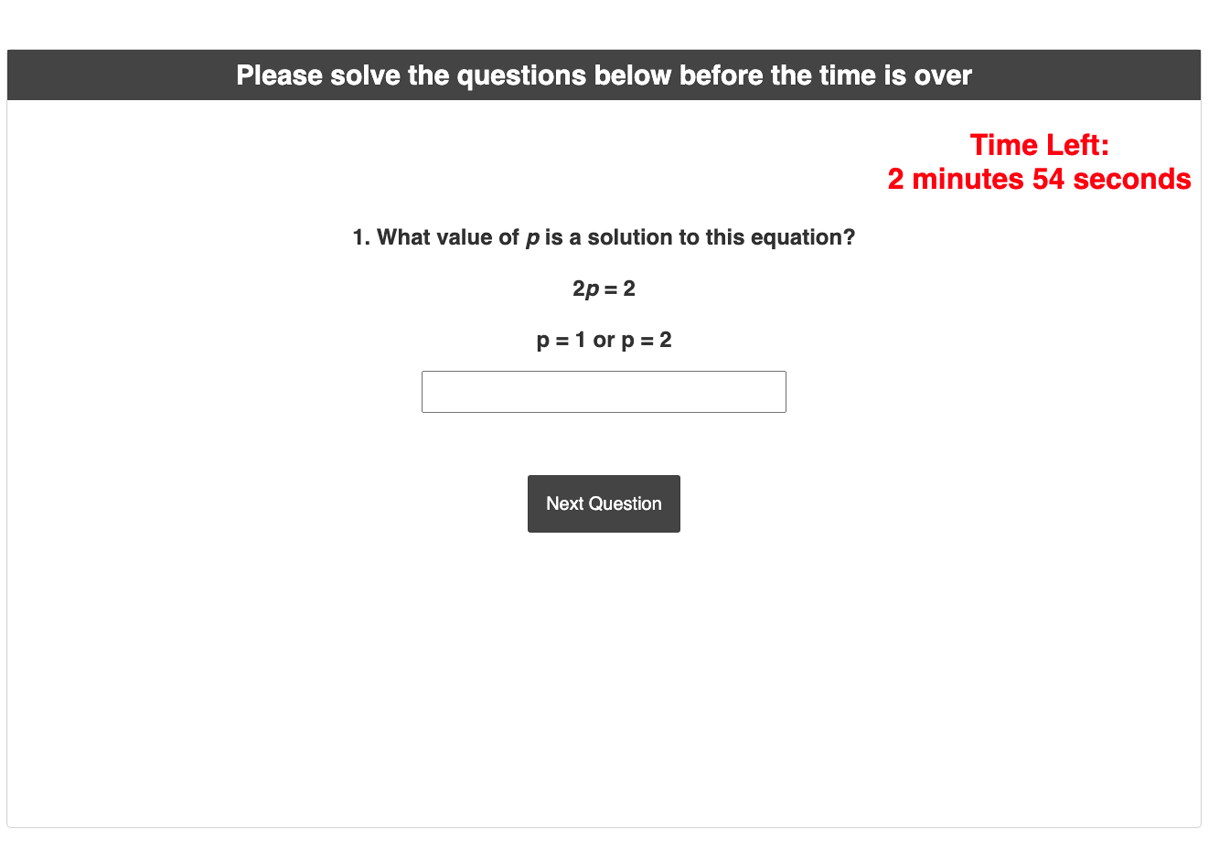}
        \caption{Screenshot of the eliciting task for the red zone, a maths quiz with a timer.}
        \label{fig:screenshot-maths}   
    \end{subfigure}%
    ~ 
    \begin{subfigure}[t]{0.5\textwidth}
        \centering
        \includegraphics[width=\textwidth]{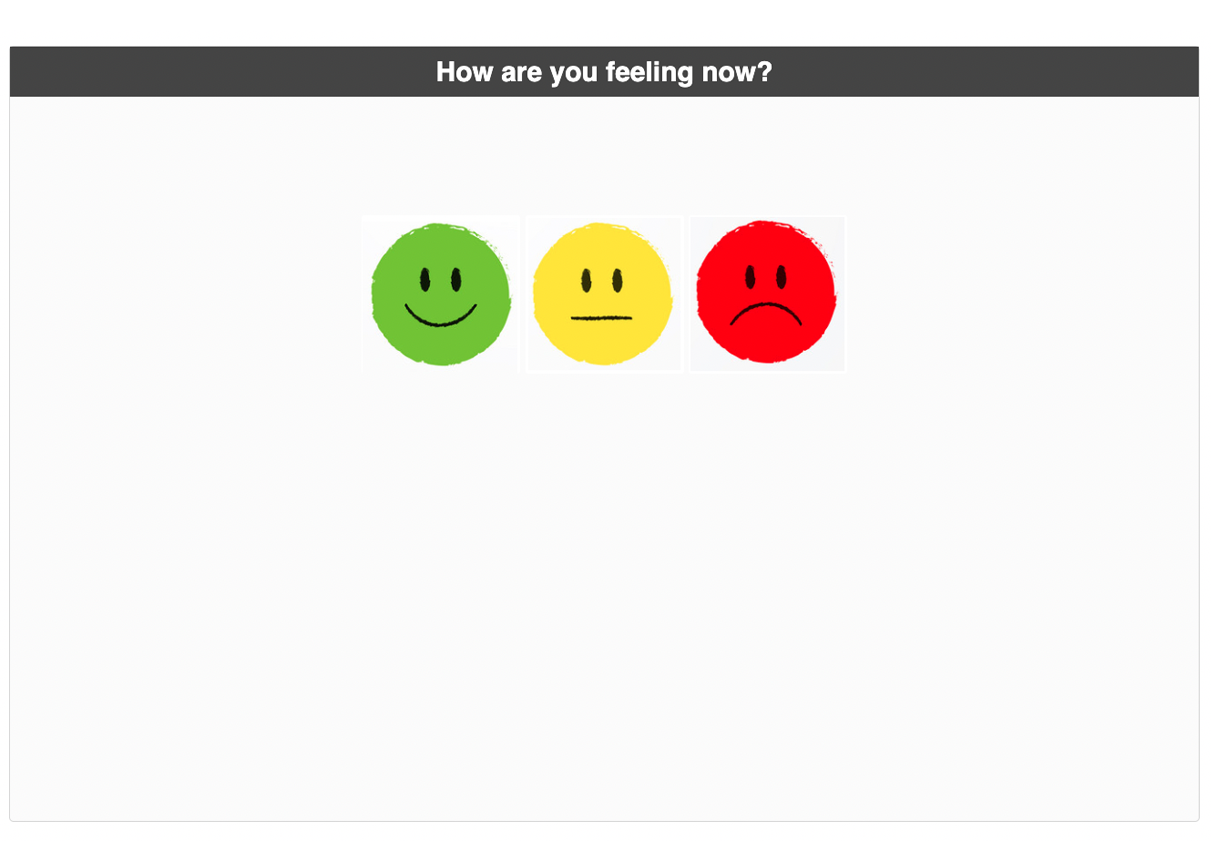}
        \caption{Screenshot of the emotion self-report to be selected by the participant after each eliciting task.}
        \label{fig:screenshot-self-report}
    \end{subfigure}

    \centering
    \begin{subfigure}[t]{0.5\textwidth}
        \centering
        \includegraphics[width=\textwidth]{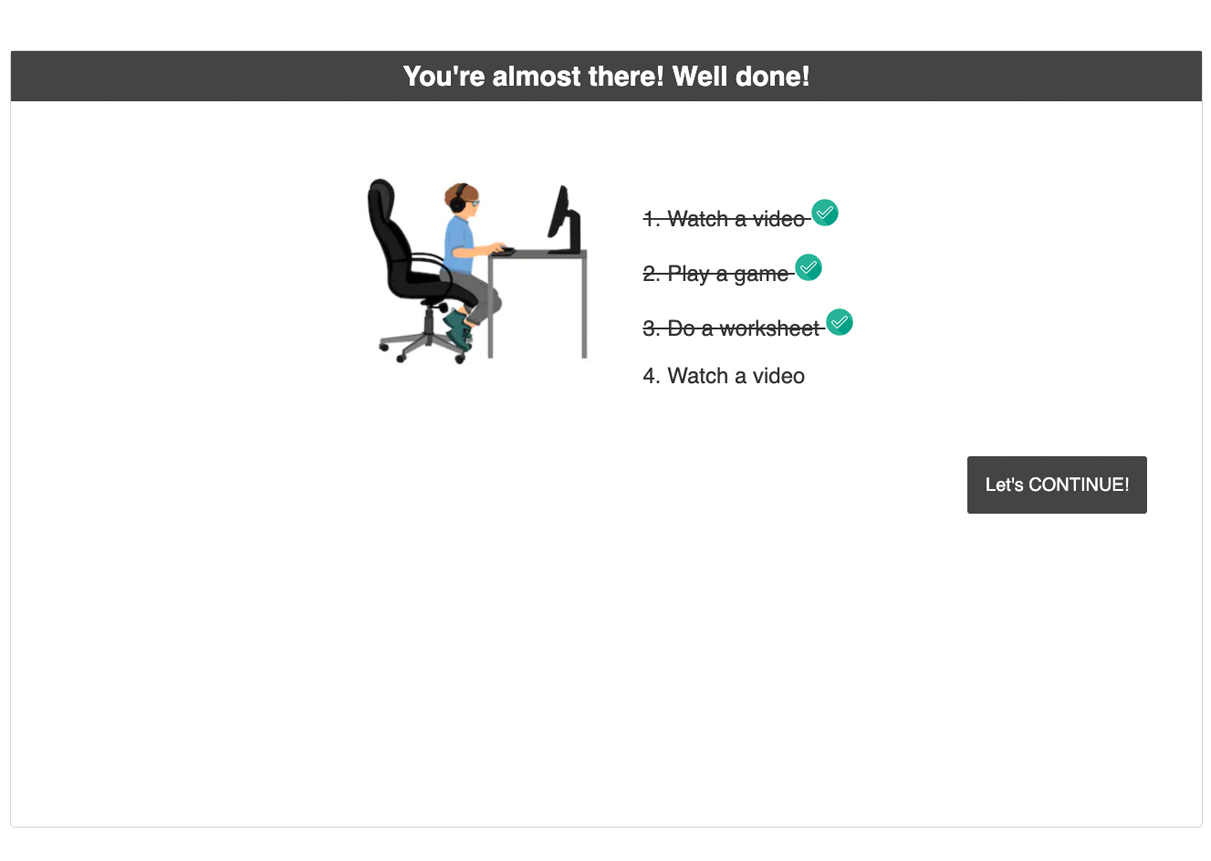}
        \caption{Screenshot of the visual agenda to support participants regarding the passing time.}
        \label{fig:screenshot-agenda}
    \end{subfigure}%
    ~ 
    \begin{subfigure}[t]{0.5\textwidth}
        \centering
        \includegraphics[width=\textwidth]{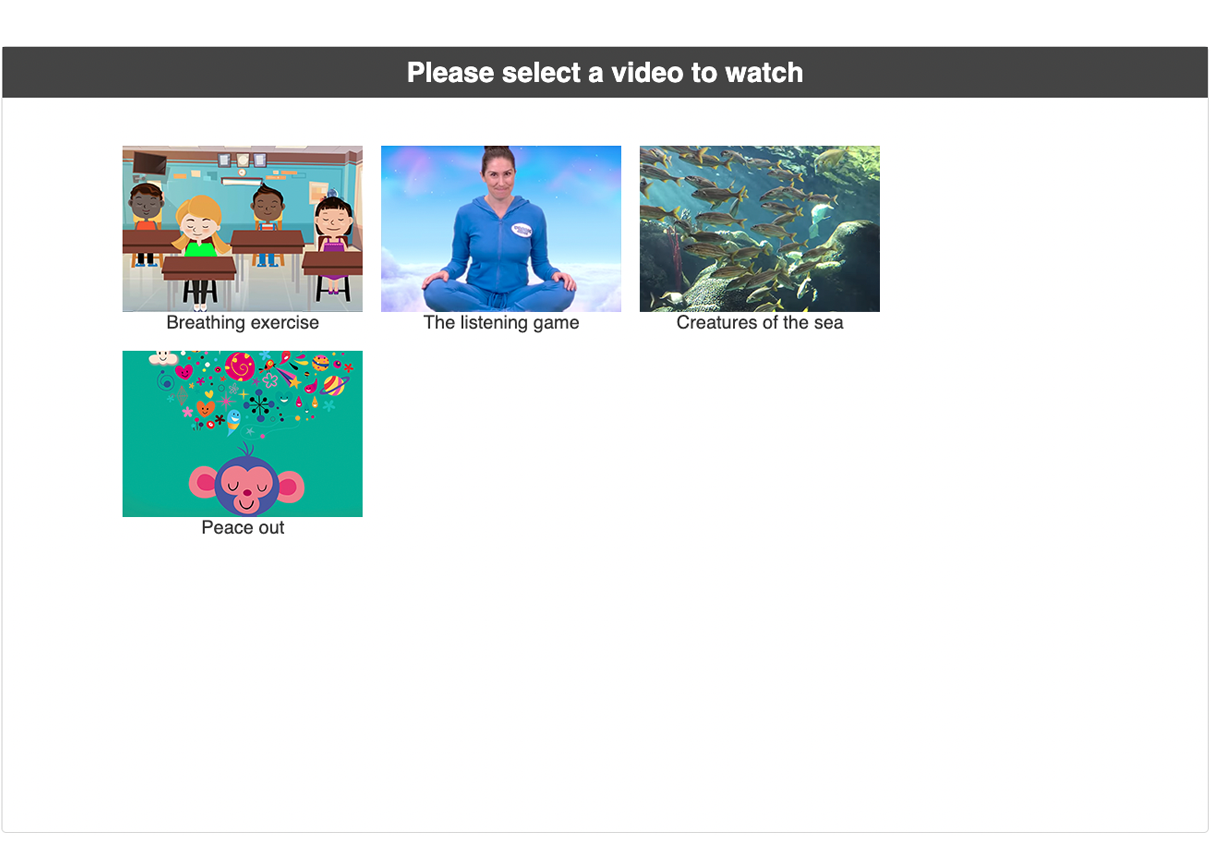}
        \caption{Screenshot of the calming activities to be selected by the participant between each eliciting task.}
        \label{fig:calming-activities}
    \end{subfigure} 
    \caption{Screenshots of the Task Environment System, a web system created to support the study session with participants. Participants accessed the system through a web browser and completed the eliciting tasks. The system can be accessed at \textit{http://task-environment.datascienceinstitute.ie/}.}
    \label{fig:task-environment-screenshots}
\end{figure*}

During the study session, the participant accessed a web-based software we created that encompasses four different eliciting tasks and calming activities. Each of the four eliciting tasks aims to evoke one of the emotion zones. Figure~\ref{fig:task-environment-screenshots} shows screenshots of some parts of the Task Environment System.

The eliciting tasks follow the sequence: green, yellow, red and blue zone. The order of the eliciting tasks focused on the participant's emotional well-being. It is expected that a participant already experiences some level of anxiety from participating in the study. They are talking to someone they do not know and doing something outside their routine, two things that tend to cause anxiety for individuals with ASD. We, therefore, did not want to aggravate it by starting the session by eliciting negative emotions. Hence, the first task is to calm the participant, eliciting the green zone, starting with a relaxed, happy and calm tone. Following the green zone, we elicit the yellow zone, which involves some uneasiness and agitation, and next, we elicit the red zone. The last emotion zone is the blue zone since we expected that by this point, the child would start to get tired, so we would naturally observe the blue zone. We defined this order to foster a more gentle progression towards the more challenging emotions so that the change to the red zone would not happen too abruptly, which could cause increased emotional discomfort. 

The eliciting tasks are as follows: watching a video with funny and cute animals for the green zone (3 minutes), playing a game for the yellow zone (3 minutes), completing a challenging maths worksheet with a visible timer for the red zone (3 minutes, see Figure~\ref{fig:screenshot-maths}), and watching a boring video for the blue zone (3 minutes). Right after each eliciting task, the system includes a prompt for a simplified emotion state self-report with three options, happy face, neutral face and unhappy face (see Figure~\ref{fig:screenshot-self-report}). The self-report does not include the four emotional zones because it could be too complex for the children, especially considering that alexithymia, difficulty in naming one's emotions, is common in individuals with autism. 

Between each of the tasks, the system also includes a calming activity (1.5 minutes) which serves a twofold purpose: First, to help the child to calm and regulate their emotions, decreasing the possibility of emotional discomfort, and second, to return the participant's emotions to a baseline before moving forward to the next emotion zone elicitation (see Figure~\ref{fig:calming-activities}). 

In addition, the system includes a visual schedule with the four tasks crossed throughout the session to help the participants have a sense of the time passing and what to expect during the session and when the session is close to the end. This technique is widely used in psychology to support children with ASD so they feel less anxious (see Figure~\ref{fig:screenshot-agenda}).

The system at \url{http://task-environment.datascienceinstitute.ie/} is configurable depending on the session number, i.e. one or two, the child's age for the maths worksheet, i.e., younger, default and older, and the time allocated for each task and activity. The system is currently configured for the first session, with ``older'' as a maths worksheet configuration and the default activity time (3 minutes per task). The source code for the system is available for the research community by request. Hence other researchers can reproduce the same study session or configure it to meet their research goals.

\section{Dataset Creation} \label{sec:dataset-creation}
\subsection{Data Annotation by Parents}
\begin{figure}[ht!]
\centering
\includegraphics[width=0.8\textwidth]{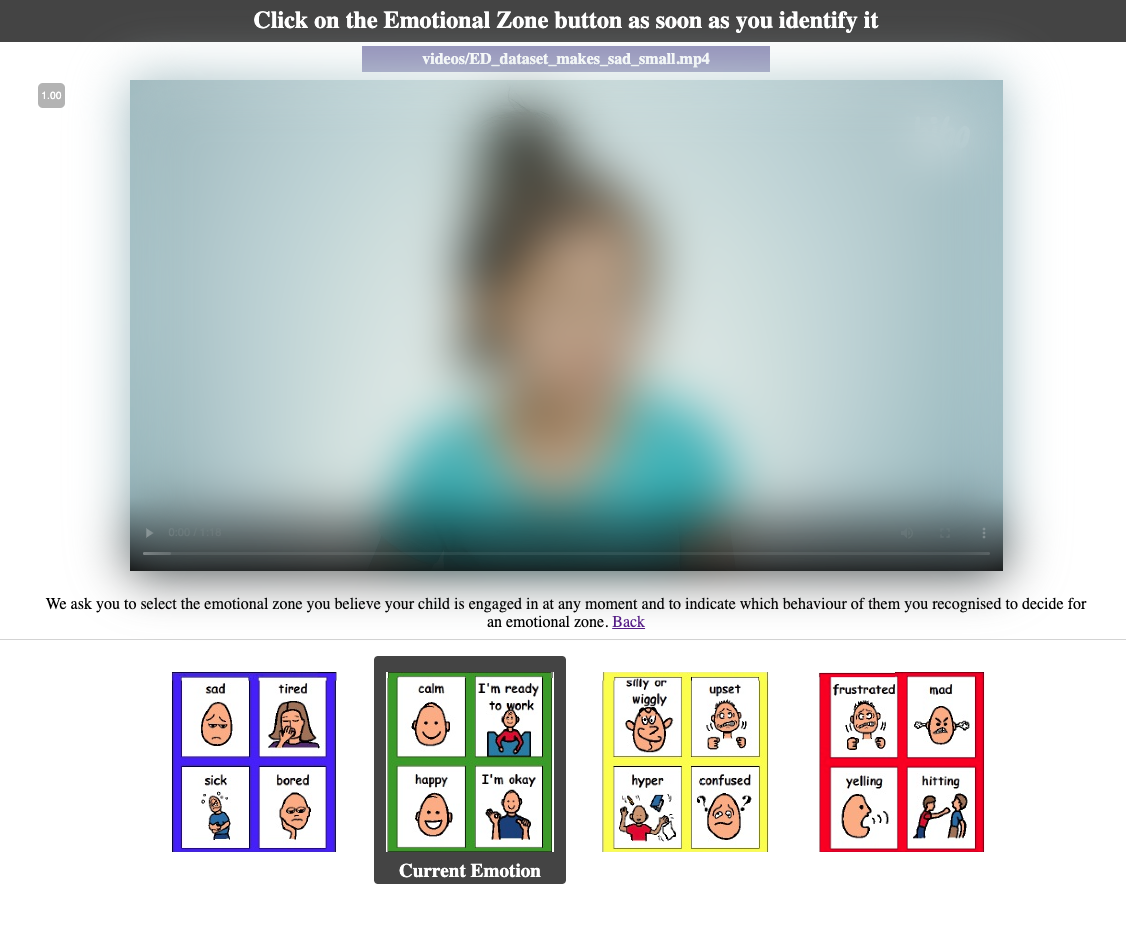}
\caption{Screenshot of the Annotation System, a web system created to support the annotation process by the participants' parents. The parent watched their child's study session recording and selected the appropriate emotion zones by clicking on the buttons below the video.} 
\label{fig:annotation-system-screenshot}
\end{figure}

Since children with autism can individually display emotions in a particular way, we asked each participant's parent to annotate their child's emotions throughout the study session, acknowledging that they know their child's emotions best. We created a system to support the annotation process. The parent's interface to the system is composed of the video recording of their child's study session with four buttons below it; each button represents an emotion zone to be clicked when the parent believes their child is displaying the given emotion zone. Figure~\ref{fig:annotation-system-screenshot} shows a screenshot of the system. Each study session recording was divided into videos of around five minutes to facilitate the annotation by the parent. We instructed the parent to watch the recording; they should click on the button representing the emotion as soon as they identified an emotion zone. At the end of the annotation, we also asked the parents to answer a brief questionnaire regarding the annotation session. None of the parents reported difficulty identifying their child's emotions during the annotation session. 

\subsection{Dataset Creation Overview}

\begin{figure}[ht!]
\centering
\includegraphics[width=\textwidth]{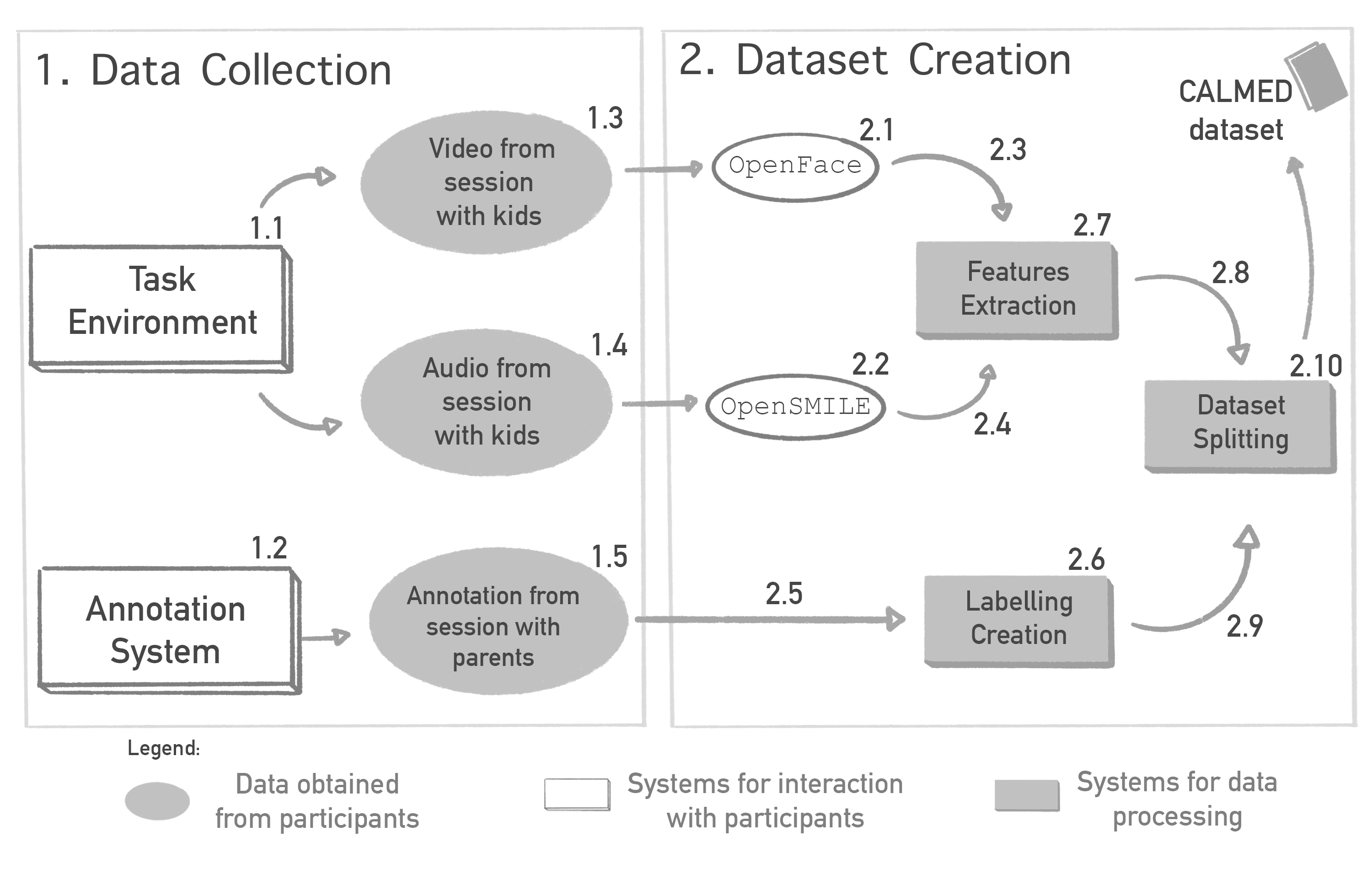}
\caption{Pipeline with an overview of the CALMED dataset creation process, from data collection to dataset creation, including all the systems developed.} 
\label{fig:pipeline-systems}
\end{figure}

The study sessions with the participants and the annotation sessions with the parents generated the raw data to be processed as a working annotated dataset. The raw files are converted into a working dataset via a pipeline process depicted in Figure~\ref{fig:pipeline-systems}. First, following the diagram in Figure~\ref{fig:pipeline-systems}, we conducted study sessions (1.1) with the participants and annotation sessions (1.2) with the parents, obtaining the data for the dataset creation (1.3-1.5). We then used OpenFace~\cite{baltrusaitis2018openface} and OpenSMILE~\cite{eyben2010opensmile} to extract video and audio features, respectively (2.1 and 2.2). We then process the result of the parent's annotation sessions (2.5) into a dataset, i.e., a .CSV file with annotation from each session (2.6). In sequence, we clean, organise and synchronise the features datasets to the annotation datasets (2.7). At this point, we have a set of .CSV files with video, audio and annotation features for each session, all containing a column with the video timestamp that we use for synchronisation (2.8 and 2.9). These .CSV files represent the whole dataset without any split into different sets. Lastly, we created a split of the train, validation and test sets, with 80/10/10 proportions (2.10), resulting in the CALMED dataset.  

The following sections present a detailed description of each of those steps.

\subsection{Video Features Extraction} \label{subsec:video-features}
Each study session produces audio and video files of around 30 minutes in length. To decrease the file size and facilitate the processing step, we split audio/video files into smaller files, each of around five minutes. 

We used OpenFace~\cite{baltrusaitis2018openface} to generate the visual features for the CALMED dataset. OpenFace is a toolkit composed of computer vision algorithms capable of executing essential tasks for visual affective computing, such as facial landmark, facial action unit detection, head pose, and eye gaze estimation. OpenFace has been widely used in the literature to generate visual features for analysing facial behaviour.

We passed the five-minute video files to the OpenFace tool for video feature extraction. The tool extracts visual features for each frame, returning a massive .CSV file with all the numerical visual features and timestamps (the output 2.3 in Figure~\ref{fig:pipeline-systems}). We use OpenFace's output as the input to a system created to clean and organise the visual features. In the \textit{Features Extraction} system, 2.7 in the Figure~\ref{fig:pipeline-systems}, we separate the visual features into groups, e.g., Facial Action Units (AU) and gaze. We also organise the output from OpenFace into our desired time window of 200 milliseconds, with the correct timestamp to be used to synchronise the visual data with the audio and annotation data. The visual features included in the resulting dataset are: Facial Action Units (FAUs), face landmarks, eye landmarks, gaze, and head pose, comprising 669 visual features in total. Table~\ref{tab:video-features} summarises the number of features per group. A comprehensive description of the OpenFace features can be found in the project's publication~\cite{baltrusaitis2018openface}. 

\begin{table}[!ht]
\centering
\caption{Video features with feature group, description and number of features.}
\label{tab:video-features}
\begin{tabular}{ccl}
\multicolumn{3}{c}{Video Features}                                                                 \\ \hline
                             & \# of Features & \multicolumn{1}{c}{Description}                    \\ \hline
AUs                          & 35             & Facial Action Units                                \\
Gaze                         & 8              & Eye gaze direction vector                          \\
2d\_eye\_landmark            & 112            & Location of 2D eye region landmarks in pixels      \\
3d\_eye\_landmark            & 168            & Location of 3D eye region landmarks in millimetres \\
face\_2d\_landmarks & 136            & Location of 2D landmarks in pixels                 \\
face\_3d\_landmarks          & 204            & location of 3D landmarks in millimetres            \\
head\_pose                   & 6              & Location of the head with respect to camera        \\ \hline
\textbf{Total}               & 669            &          \\ \hline                                          
\end{tabular}
\end{table}
\subsection{Audio Features Extraction} \label{subsec:audio-features}

We similarly repeated the steps described in the last section for audio features extraction using OpenSMILE~\cite{eyben2010opensmile} for acoustic features creation. OpenSMILE is a robust system capable of extracting a vast set of audio features. We defined a selection of audio features to be included in the dataset based on the Extended Geneva Minimalistic Acoustic Parameter Set (eGeMAPS). The eGeMAPS is a set of features curated as a recommendation of audio parameters for voice research and affective computing. The parameters were selected based on: 1)  how extensively and successfully these parameters were used in the literature, 2) the theoretical significance of the parameter, and 3) the potential of an acoustic parameter to represent physiological changes that occur in the voice during affective processes~\cite{eyben2015geneva}. From eGeMAPS, we selected 75 parameters resulting from Low-Level Descriptors (LLDs) extracted from audio files and sorted them into four groups: frequency, energy/amplitude, spectral (balance) parameters, and additional temporal features. Table~\ref{tab:audio-features} summarises the number of features per type. 

\begin{table}[!ht]
\centering
\caption{Audio Features with feature group, description and number of features. }
\label{tab:audio-features}
\begin{tabular}{ccc}
\multicolumn{3}{c}{Audio Features}                                                                                                                                      \\ \hline
                  & \multicolumn{1}{c}{\# of Features} & \multicolumn{1}{c}{Description}                                                                                \\ \hline
Frequency         & 24                                 & \begin{tabular}[c]{@{}l@{}}Frequency related parameters, \\ e.g., pitch, jitter.\end{tabular}                  \\
Energy Amplitude  & 15                                 & \begin{tabular}[c]{@{}l@{}}Energy/Amplitude related parameters, \\ e.g., shimmer, loudness.\end{tabular}       \\
Spectral Balance  & 30                                 & \begin{tabular}[c]{@{}l@{}}Spectral (balance) parameters, \\ e.g., alpha ration, Hammarberg Index\end{tabular} \\
Temporal Features & 6                                  & \begin{tabular}[c]{@{}l@{}}Temporal features related, \\ e.g., rate of loudness peaks.\end{tabular}            \\ \hline
\textbf{Total}    & 75                                 &   \\    \hline                                                                                                        
\end{tabular}
\end{table}

The methodology described above to create and synchronise the dataset's visual and audio features is based on the methodology used by AVEC-2018~\cite{ringeval2018avec} when using the RECOLA~\cite{ringeval2013introducing} multimodal dataset, which shares some similarities with the CALMED dataset, including the video and audio data inputs. 

\subsection{Dataset Labelling}
From the annotation sessions, we get a .sqlite file with each emotion zone selected by the annotator and the video's timestamp indicating when the emotion was chosen. From this file, we produce a .csv file where each row represents a time window of 200 milliseconds and the assigned label for the given time window. Each row corresponds to one example within the labels dataset. 

The steps to create the labels dataset are as follows, they are executed in the \textit{Labelling creation} system (marked as 2.6 in Figure~\ref{fig:pipeline-systems}): 
\begin{enumerate}
\item Create a .CSV file with a row representing each time window of 200 milliseconds of a given video file of a session with a participant;   
\item Read the emotion zones selected by the annotations in the .sqlite file and;
\item Assign the corresponding annotation for each row in the dataset.
\end{enumerate}

One challenge during this step was identifying how close the annotation time was to the real emotion utterance. During the annotation, we collect the time of the video when the annotator pressed the emotion button, which occurs with an inherent delay after the parent identifies the emotion. In other words, the emotion expression started some point before the time the parent marks it. For this study, we empirically defined the delay time as equal to one second. In practice, we marked in the dataset that the emotion started one second before the timestamp selected by the annotator. 

\subsection{Dataset Splitting}\label{sec:labelling}
After creating a features dataset of audio, video and annotation labels,  we then split it into three different sets: train, validation and test sets, with a proportion of 80/10/10, respectively, each of them containing the same four emotion zones distribution. To accomplish the same distribution of emotion within train, validation and test sets, we followed the steps described below. We utilised the sample method of Pandas Python library with a random state value of 25. Each participant's data appear in the train, validation and test sets, i.e., there are no participant data only present in one of the splitting sets. 

Within a given session's data: 

\begin{enumerate}
\item We selected all the examples of a given emotion and split them randomly into train, validation and test with the proportion of 80/10/10, respectively,  
\item We repeated the previous step with the other emotions,  
\item We then concatenate all the same dataset types together, i.e., the train data from green, yellow, blue and red emotion zones, and so on, generating then train, validation and test set following the same distribution of labels. 
\end{enumerate}

The splitting method described above considers each time window of 200ms as an independent example in the dataset, not following the time sequence of the video. However, CALMED dataset includes all video/audio/labels timestamps, allowing researchers to create alternative splits according to their research needs. Any additional split type created by the research team will be available under request. 

\section{Analysis and Discussion} \label{sec:analysis-discussion}
\subsection{Dataset in Numbers}

The CALMED dataset comprises a total of 57,012 annotated examples, each representing a time window of 200 milliseconds, created from data collected across all the participant sessions. The number of examples per class is as follows: 30,882 examples of green, 9,858 examples of yellow, 3,179 of red, and 13,093 of blue. When looking at the splitting numbers, we have 45,608 examples in train, 5,702 in validation and 5,702 in test sets. Table~\ref{tab:dataset-numbers} summarises the number of examples per class in the different sets of the data. 

The number of examples in the green emotion zone is significantly larger than in the red emotion zone, i.e. around ten times more. This unbalanced dataset characteristic was expected due to our decision to prioritise the participant's well-being. Since the red class models some negative and demanding emotions, during the sessions, we wanted to elicit the red zone in the least amount of time so as not to cause emotional harm to the participants. Thus, the resulting dataset presents a significantly lower number of examples of the red class than the green class.

\begin{table}[h!]
\caption{Number of examples per class across the different sets of the data.}
\label{tab:dataset-numbers}
\centering
\resizebox{0.7\columnwidth}{!}{%
\begin{tabular}{l|llll|l}
\hline
            & \textbf{Green   } & \textbf{Yellow  } & \textbf{Red       } & \textbf{Blue      } & \textbf{Total} \\ \hline
\textbf{Whole Dataset  }   & 30,882         & 9,858           & 3,179        & 13,093        & \textbf{57,012}         \\ \hline
Train Set     & 24,704         & 7,886           & 2,543        & 10,475        & 45,608         \\
Dev Set       & 3,089          & 985             & 319          & 1,309         & 5,702          \\
Test Set      & 3,089          & 987             & 317          & 1,309         & 5,702          \\ \hline
Participant 1 & 5,158          & 788             & 390          & 3,156         & 9,492          \\
Session 1.1   & 5,158          & 788             & 390          & 3,156         & 9,492          \\ \hline
Participant 2 & 8,139          & 3,788           & 1482         & 3,020         & 16,429         \\
Session 2.1   & 3,664          & 2,444           & 829          & 1,386         & 8,323          \\
Session 2.2   & 4,475          & 1,344           & 653          & 1,634         & 8,106          \\ \hline
Participant 3 & 8,534          & 3,193           & 595          & 4,456         & 16,778         \\
Session 3.1   & 4,939          & 1,072           & 516          & 2,265         & 8,792          \\
Session 3.2   & 3,595          & 2,121           & 79           & 2,191         & 7,986          \\ \hline
Participant 4 & 9,051          & 2,089           & 714          & 2,461         & 14,315         \\
Session 4.1   & 4,483          & 800             & 278          & 1,474         & 7,035          \\
Session 4.2   & 4,568          & 1,289           & 436          & 987           & 7,280          \\ \hline
\end{tabular}%
}
\end{table}

\subsection{Applicability, Limitations and Perspectives}
The CALMED dataset was created as part of a research project aiming to create a multimodal emotion detection system for children with ASD. Thus, we foresee that the main scenario in which CALMED can be applied is to train and test machine learning systems for affective computing tailored to children with ASD. Insights and knowledge are expected to emerge from creating a classifier model, such as, the most viable features, modalities and fusion layer techniques   to be applied for emotion detection in the context of people with ASD. Moreover, this has the potential to generate knowledge for both the affective computing and psychology fields. 

There are other subsets of the general population that are known to also express emotions differently from a typical population, e.g. people with schizophrenia, brain damage, and some mental health condition such as depression. All those populations could also benefit from the advancement of affective computing systems. However, to accomplish that, it would also be necessary to create annotated datasets featuring these populations. The methods described, and the systems shared in this paper could be extended, adapted and applied to designing and creating these annotated datasets. 

The CALMED dataset could also be applied to evaluate the viability of creating a person-independent model, i.e. an emotion detection model applied to individuals whose data was not present in the training dataset. For this purpose, the dataset would need to split so that some participants' data do not appear in the training dataset, which is not the case now, as mentioned in Section~\ref{sec:labelling}.

Sharing only the extracted features as opposed to the original files can be considered a limitation since other researchers will not be able to extract their features. However, that was a middle-ground solution to both protect the participants' privacy and have a first-of-its-kind valuable resource shared within the research community. 

As reported by previous works, the research team faced the same challenges in recruiting children as participants for creating a dataset. Thus, one of the main limitations of the dataset is the limited number of subjects, which implies that it does not encompass a wide diversity in the extracted data. However, we were able to extract meaningful knowledge from the data collected and expect to have more participants in the future to generate more data. 

Another aspect to note concerns the annotation performed by participants' parents. Some of the previous works decided to use annotation by parents, while others selected annotation by specialists instead. We do not yet know if the parents' annotation carries any biases or noise or to what extent they are reliable, so it would be helpful to have an additional annotation from a specialist in ASD who does not know the participants and compare these two sets of annotation—checking for agreement rate, differences in the distribution of emotions, among others.

\section{Conclusion and Future Work}

This paper introduced CALMED (Children, Autism, Multimodal, Emotion, Detection), a multimodal (video-audio) emotion detection dataset featuring children aged 8-12 years old with a level 1 diagnosis of autism. The multimodal dataset includes audio and video features extracted from recording files of study sessions with participants, together with annotation provided by their parents into four target classes (green, yellow, red and blue) based on the ``the zones of regulation'' framework~\cite{kuypers2013zones}. The generated dataset includes a total of 57,012 examples. Each example represents a time window of 200ms (0.2s) of audio and video extracted features and an emotion annotation. The resulting dataset is available for the research community upon request. 

This paper's contribution is unique since it shares an audio/video features dataset with the research community. By sharing the features dataset with the research community, we maintain the privacy of the subjects while still making available resources to facilitate further research in this emotion detection application, which has the potential to improve systems of automatic ED applied to children with ASD. We hope that our experience and methods described here contribute to future research applications of affective computing in this scenario, for instance, educational platforms, and self-identify emotions assistance tools, among others. 

The small sample size of participants did not allow us to create a more diverse dataset regarding gender, culture, and language. Thus, further participant recruitment and data collection would benefit the dataset, including children from other cultures, countries and backgrounds. Using the task environment and annotation systems shared here, the data collected would be consistent, and other research groups could further expand the dataset. 

A further study will investigate how a specialist annotation compares with the parents' annotation of the participants' emotions. Our research group has already completed the annotation of emotion zones by such a specialist. It is now working on generating a new dataset which will be compared with the original annotation by the parents. 

A natural progression of this work is to create a multimodal ED machine learning system, an ongoing work in our research group. Initial results have been obtained, and further steps are being taken to improve the classification accuracy. The resulting system will feature in a future publication.

\section{Acknowledgements}
The research team would like to thank Dr Ciara Gunning who provided specialised advice, training on how to interact with children with ASD, and revision of the data collection experiment design and materials, together with support on the recruitment of participants. We also thank Aindrias Cullen for his comprehensive advice on data protection legislation, so we could design a project that is compliant with GDPR. We thank still Adhara Correa Soto, Alisha Garvey, Hannah Callanan, and Liam Finnerty for their participation and support during study sessions with participants. And lastly, we thank all the lovely participants and their parents who made this project possible. We hope you harvest the results of this research. It was a pleasure to meet and work with each of you. 

This publication has emanated from research conducted with the financial support of Science Foundation Ireland under Grant number SFI/12/RC/2289\_P2\\\\\\\\

%
%
%
\bibliographystyle{splncs04}
\bibliography{references}

\end{document}